\documentclass[letterpaper]{article} 
\usepackage{aaai2026}  
\usepackage{times}  
\usepackage{helvet}  
\usepackage{courier}  
\usepackage[hyphens]{url}  
\usepackage{graphicx} 
\usepackage{natbib}  
\usepackage{caption} 
\usepackage{float}
\usepackage{amsmath}
\usepackage{xspace}
\usepackage{amsthm} 
\usepackage{amssymb}
\usepackage{subcaption}
\usepackage{amsfonts}
\usepackage{tcolorbox}
\usepackage{algorithm}
\usepackage{algorithmic}
\newtheorem{theorem}{Theorem}

\newtheorem{proposition}{Proposition}
\newtheorem{definition}{Definition}
\newtheorem{lemma}{Lemma}
\newtheorem{remark}{Remark}
\newtheorem{problem}{Problem}
\newtheorem{assumption}{Assumption}

\newcommand{\R}{{\mathbb{R}}}

\newcommand{\E}{{\mathbb{E}}}
\newcommand{\C}{{\mathcal{C}}}
\newcommand{\norm}[1]{\left\| #1 \right\|_2}
\newcommand{\normm}[1]{\left\| #1 \right\|}

\usepackage{newfloat}
\usepackage{listings}
\DeclareCaptionStyle{ruled}{labelfont=normalfont,labelsep=colon,strut=off} 
\floatstyle{ruled}
\newfloat{listing}{tb}{lst}{}
\floatname{listing}{Listing}
\title{Just Few States are Enough: Randomized Sparse Feedback for Stability of Dynamical Systems}
\author{    
    Zaid Hadach,
    Hajar El Hammouti,
    El Houcine Bergou,
    Adnane Saoud
}
\affiliations{
    College of Computing, Mohammed VI Polytechnic University (UM6P)\\
    Ben Guerir, Morocco\\
    \{zaid.hadach, hajar.elhammouti, elhoucine.bergou, adnane.saoud\}@um6p.ma
}

\begin{document}

\maketitle

\begin{abstract}
While classical control theory assumes that the controller has access to measurements of the entire state (or output) at every time instant, this paper investigates a setting where the feedback controller can only access a randomly selected subset of the state vector at each time step. Due to the random sparsification that selects only a subset of the state components at each step, we analyze the stability of the closed-loop system in terms of Asymptotic Mean-Square Stability (AMSS), which ensures that the system state converges to zero in the mean-square sense. We consider the problem of designing both a feedback gain matrix and a measurement sparsification strategy that minimizes the number of state components required for feedback, while ensuring AMSS of the closed-loop system. Interestingly, (1) we provide conditions on the dynamics of the system under which it is possible to find a sparsification strategy, and (2) we propose a Linear Matrix Inequality (LMI) based algorithm that jointly computes a stabilizing gain matrix, and a randomized sparsification strategy that minimizes the expected number of measured state coordinates while preserving the AMSS. Our approach is then extended to the case where the sparsification probabilities vary across the state components. Based on these theoretical findings, we propose an algorithmic procedure to compute the vector of sparsification parameters, along with the corresponding feedback gain matrix. To the best of our knowledge, this is the first study to investigate the stability properties of control systems that rely solely on randomly selected state measurements. Numerical simulations demonstrate that, in some settings, the system achieves comparable performance to full-state feedback while requiring measurements from only $0.3\%$ of the state coordinates.
\end{abstract}

\section{Introduction}
In control theory, the primary objective is to design control strategies that ensure a system behaves according to a desired specification \cite{aastrom2021feedback}. Among the most fundamental specifications is the requirement for the system to converge to a specific operating point, most commonly the origin \cite{khalil2009lyapunov}. This property, known as stabilization to the origin, is essential in many applications where the system must settle into a rest state or nominal condition after initial transients \cite{hespanha2018linear}. Achieving this goal typically involves the design of a feedback controller that adjusts the input based on the system's current state, ensuring asymptotic convergence to the origin. 

In closed-loop control systems, feedback is critical for ensuring asymptotic stability, but measuring the full state vector can be resource-intensive, particularly in high-dimensional systems or constrained environments. To address this, we introduce in this paper a novel control paradigm in which the controller accesses only a randomly selected subset of the state vector at each time step. A clear motivation for incorporating sparsity in control design stems from recent technological advances that have enabled individual components of large-scale systems to integrate their sensing, actuation, communication, computation, and decision-making capabilities. 

In this paper, we study the AMSS of linear discrete-time systems under a feedback control strategy with access to a randomized sparse state vector. Our main objective is to minimize the expected sparsity, the average number of active sensors, while ensuring AMSS. To this end, we introduce randomized sparsification strategies and analyze the system's expected dynamics. We show that the expected state converges to zero if the closed-loop system is deterministically asymptotically stable. We then formally define expected sparsity and study the system's variance, with a focus on stabilizing the system using the minimum expected number of sensors. We derive conditions under which AMSS can be achieved and propose algorithms that jointly design the feedback controller along with the sparsification probabilities minimizing the expected sparsity. Finally, we validate our results through experiments. Proofs are provided in Appendix A, and additional experiments in Appendix B.


 Our contributions are summarized as follows:
\begin{itemize}
    \item \textbf{First study of randomized sparse-feedback control:} Providing sufficient conditions for the system dynamics to support randomized sparsification and integrating these conditions into controller design that ensures minimum expected sparsity while achieving AMSS of the system.
    \item \textbf{Extension to adaptive sparsification:} Tailored sparsification strategies for each state vector component to enable flexible adjustment of sparsity based on the difficulty of operating or measuring the state component.
    \item \textbf{Efficiency and practicality:} Experimental results underscore the practical impact of this work, demonstrating that the proposed approach achieves performance comparable to full-state feedback while requiring measurements from only $0.3\%$ of state coordinates in some settings. This drastic reduction in measurement requirements is a game-changer for large-scale systems, where sensing all state variables is often infeasible due to cost, energy, or bandwidth limitations. 
\end{itemize}
\section{Related work}

\textbf{Asymptotic stability via deterministically-sparse feedback controllers.} 
 Numerous works deal with the sparsity of the control gain matrix to reduce sensor/actuator usage and communication bandwidth, which applies to networked control systems or large-scale systems. For instance, in \cite{bykov2018sparse}, the authors focused on sparsity in the control gain matrix (zero rows in $K$) to reduce the number of control inputs. Another way to achieve sparsity deterministically is by minimizing the $l_1$ norm, which leads to a gain matrix $K$ with zero columns or rows \cite{lin2012sparse}, \cite{polyak2014sparse}. The sparsity structure is fixed and designed a priori, see \cite{polyak2013lmi} for details. Furthermore, in \cite{sun2025row}, the authors introduced a mixed-integer programming approach to address row cardinality constraints, in contrast to the $l_1$ norm-based convexification method. However, all these approaches mainly focus on deterministic sparsity. Moreover, in \cite{9928443}, the authors provide sufficient and necessary conditions for achieving stability using $s$-sparse control inputs. Their work only deals with deterministic sparsity under open-loop control strategies where the control inputs depend only on the initial point and parameters of the system. As stated in \cite{bishop2011modern}, open-loop control strategies, lacking feedback, can not adapt to disturbances or parameter variations, resulting in unreliable performance. Our work presents an analysis of randomized sparsity and deals with the stability of closed-loop systems.
\\
\textbf{Sparsity in distributed learning}.  In the context of limited access to the full state, \cite{cheng2021data} addresses efficient data representation to maintain control performance by focusing on compressing high-dimensional time series data, such as cellular demand or electricity load, into a low-dimensional representation to reduce communication over a bandwidth-constrained network. Our work is also inspired by gradient vector sparsification used in the context of distributed machine learning. For example, in \cite{wangni2018gradient}, the authors introduce a sparsification technique that minimizes communication costs while keeping the increase in gradient variance small, ensuring that the optimization algorithm converges with only a slight increase in iterations. 
\\
{\textbf{Sensor scheduling.} The existing literature on sensor scheduling can be grouped into three main directions: (1) Several approaches \cite{maity2022sensor,vitus2012efficient} analyze randomness in settings where sensors communicate over constrained channels to remote estimators, focusing on estimation performance only and not on controller synthesis. (2) Other works, such as \cite{kundu2019stabilizing}, examine scenarios with multiple systems sharing a communication channel, where periodic scheduling is combined with randomness arising from network-induced losses. (3) Continuous-time co-design approaches such as \cite{nugroho2019algorithms} provide time-invariant deterministic schedules.
To the best of our knowledge, the proposed approach in our paper is the first framework that jointly designs time-varying randomized sensor selection procedures together with the control gain $K$ to ensure global mean-square asymptotic stability.}
\section{Preliminaries}
\subsection{Mathematical Notations}
We denote by $\norm{.}$ the two norm on $\R^n$. The $n$-by-$n$ identity matrix is denoted by $I_n$. Given $A\in \R^{n\times n}$, we denote by  \( \sigma(A) \) the set of all eigenvalues of $A$, by $\rho(A):= \max \{ |\lambda| : \lambda \in \sigma(A) \} $ the spectral radius of $A$, by $\normm{.}$ the spectral norm of $A$ given by: $\normm{A}:=~\text{sup}_{\norm{x}=1} \norm{Ax}$, by $A^\top$ the transpose of $A$, by $A_{:,i}$ the $i-th$ column vector of $A$, and by $Diag(A)$ a diagonal matrix whose diagonal entries are the diagonal elements of $A$. For simplification, we denote also by $Diag(\alpha_1,\ldots,\alpha_n)$ a diagonal matrix of order $n$ with diagonal coefficients $\alpha_1,\ldots,\alpha_n$. Given $A,B\in \R^{n\times m}\times \R^{n\times m}$, we denote by $\mathcal{K}(A,B)$ the set of stabilizing control gain matrices defined by $\mathcal{K}(A,B) :=\{K \in \R^{m\times n} \mid \rho( A+BK)<1 \} $. We denote by $Ber(p)$ a Bernoulli random variable of parameter $p\in (0,1]$ where $p$ represents the probability of success.
The notation $A>B$ is equivalent to the matrix $A-B$ being positive definite.
\subsection{Randomized Sparsification Strategies}
Consider a discrete-time control systems $\Sigma$ of the form:
\begin{equation} 
\label{eqn:sys}
\Sigma\!:
x(k+1)=A x(k)+B u(k) 
\end{equation}
where $x\in \R^{n}$ is a state, $u\in \R^{m}$ is a control input, and $A,B$ are matrices with appropriate dimensions. In the sequel, we assume that the system is stabilizable, i.e, there exists a feedback gain matrix \( K \in \mathbb{R}^{m\times n} \) such that $\rho(A+BK)<1$, \cite{grune2000homogeneous}. 
In this paper, we consider the case where, at each time step $k$, the controller has only access to a randomly selected portion $\hat{x}(k)$ of the full state $x(k)$, i.e the control inputs are of the form $u(k) = K\hat{x}(k)$. More specifically, at each time step k, \( \hat{x}(k) \) contains a random subset of the coordinates of \( x(k) \), and the remaining coordinates are set to zero. The portion of the state accessed is selected randomly according to a time-varying map \( \mathcal{C}: \mathbb{N} \rightarrow \{0,1\}^{n \times n} \), which is a diagonal matrix of the form \( k\mapsto \text{Diag}(c_1(k), \dots, c_n(k)) \), where \( (c_i(k), i\in \{1,\ldots,n\})\) are independent random variables, each determines whether the corresponding coordinate \( x_i(k) \) in $x(k)$ is included in \( \hat{x}(k) \) or set to zero at time $k$. More precisely, we randomly drop out the $ i$-th coordinate with a probability of $1-p_i$, which means the coordinate remains non-zero with a probability of $p_i\in (0,1]$. To preserve the unbiasedness of the sparsified state vector $\hat{x}(k)$, each non-zero coordinate $x_i(k)$ is scaled by $\frac{1}{p_i}$, resulting in $\frac{x_i(k)}{p_i}$. Thus, our sparsification matrix at each time step $k$ becomes: $\mathcal{C}(k)= Diag\left(\frac{c_1(k)}{p_1}, \ldots,\frac{c_n(k)}{p_n}\right)$, and $c_1(k),\ldots,c_n(k)$ are random Bernoulli variables of parameter $p_1,\ldots,p_n$ respectively.  In the rest of the paper, we will analyze the stability properties of the closed-loop system $\Sigma_{cl}$ consisting of the discrete-time control system $\Sigma$ in Eq.\ref{eqn:sys} under the feedback controller $k\mapsto u(k)=K\mathcal{C}(k)x(k)$, for a gain $K \subseteq \mathbb{R}^{n \times m}$ and the sparsification matrix $k \mapsto \mathcal{C}(k)$ defined above. The dynamics of the closed-loop system $\Sigma_{cl}$ is given by:
\begin{equation}
\label{eqn:sys_cl2}
x(k+1)=(A+BK\mathcal{C}(k))x(k) , k \in \mathbb{N}.
\end{equation}
This random sparsification aims to reduce financial costs (equipment, maintenance) and computational costs (processing time, energy) incurred from measuring the system's state vector (e.g., position, velocity) for computing control inputs, which depend on sensors, data acquisition, and processing \cite{zhong2024sparse}. As such, in this paper, we study two sparsification scenarios:
\begin{enumerate}
    \item The case where $c_i(k)$ are $Ber(p)$ where $p\in (0,1]$, and 
    \begin{equation}{\label{compressor}}
        \mathcal{C}(k)= Diag\left(\frac{c_1(k)}{p}, \ldots,\frac{c_n(k)}{p}\right).
    \end{equation}
    \item The adaptive sparsification case where  $c_i(k)$ are $Ber(p_i)$,  and the sparsification matrix is given by: 
    \begin{equation}{\label{adapcompressor}}
        \mathcal{C}(k)= Diag\left(\frac{c_1(k)}{p_1}, \ldots,\frac{c_n(k)}{p_n}\right).
    \end{equation}
\end{enumerate} 
This second scenario allows for more flexibly. It supports the prioritization of sensor monitoring of dynamically significant states while also reducing reliance on states that are costly or difficult to measure.
\subsection{Asymptotic Stability of the Expected State Vector}
In this section, we demonstrate that for any sparsification strategy described in Eq.\ref{compressor} or Eq.\ref{adapcompressor}, if there exists $ K \in \mathcal{K}(A,B) $, the expected state trajectory converges to zero regardless of the initial condition of the system. This preliminary finding represents an initial step toward achieving AMSS.
\begin{lemma}{\label{expected_stability}}
    Consider the discrete-time control system $\Sigma$ in Eq.\ref{eqn:sys} under the controller $k\mapsto u(k)=K\mathcal{C}(k)x(k)$, for a gain $K \subseteq \mathbb{R}^{n \times m}$ and where the compressor $k \mapsto \mathcal{C}(k)$ is defined by  $\mathcal{C}(k)= Diag\left(\frac{c_1(k)}{p_1}, \ldots,\frac{c_n(k)}{p_n}\right)$, where for each time step $k$, $c_1(k),\ldots,c_n(k)$ are independent Bernoulli random variable of parameter $p_1,\ldots,p_n$ respectively. The following holds
    \begin{equation}
        \mathbb{E}(x(k+1))=(A+ BK) \mathbb{E}(x(k)).
    \end{equation}
    Moreover, if $K \in \mathcal{K}(A,B)$, then the system is asymptotically stable in expectation. 
\end{lemma}
Intuitively, the proposed result states that the chosen Bernoulli-based sparsification policy ensures the asymptotic stability (in expectation) of the controlled system  for any parameters $p_1,\ldots,p_n \in  (0,1]$, if there exists a matrix $ K \in \mathcal{K}(A,B)$. However, since convergence in expectation does not imply AMSS, we further analyze the second-order moment by studying the variance to ensure the system's asymptotic stability in the mean-square sense.
\section{Problem Formulation}
It is evident that maintaining the Bernoulli parameter sufficiently close to zero is essential for implementing an effective sparsification strategy. However, sparsification inherently introduces errors by reducing the controller's influence on the system, thereby creating a trade-off that must be carefully managed. In the sequel, we provide formal definitions of expected sparsity and  AMSS, and formally outline the problem under consideration.
\begin{definition} 
Consider the discrete-time closed-loop control system $\Sigma_{cl}$ in Eq.\ref{eqn:sys}. The system $\Sigma_{cl}$ is said to be AMSS if 
$\lim _{k \rightarrow \infty} \mathbb{E}\left\|x(k)\right\|_2^2=0$ holds for any initial condition $x(0)\in \R^n$ \cite{shaikhet1997necessary}.
\end{definition}
Intuitively, the AMSS means that the system is robust and will asymptotically converge to zero while rejecting the effect of randomization on the state norm.
\begin{definition}
Consider the discrete-time closed-loop control system $\Sigma_{cl}$ in Eq.\ref{eqn:sys_cl2}. We define the expected sparsity of the system $\Sigma_{cl}$, denoted by $ES$, as the expected number of used sensors, which is formally defined as follows:
    \begin{enumerate}
        \item For scenario 1 in Eq.\ref{compressor}: $ES= p\times n.$
        \item For scenario 2 in Eq.\ref{adapcompressor}: $ES = \sum_{i=1}^{n} w_ip_i$  where $w_1,\ldots,w_n$ are weights representing the measurement difficulty or sensing constraints tied to each component of the state vector.
    \end{enumerate}
\end{definition}
Sparsity, in this context, refers to how many sensors in a system of $ n $ total sensors are expected to be {``active''} at each time step $k$. With these definitions established, we are now fully prepared to present and analyze the problem at hand.
\begin{tcolorbox}[
  colback=gray!15, colframe=black, boxrule=0.5pt, width=0.47\textwidth,height=0.13\textwidth]
  \begin{problem}
Find a sparsification strategy that minimizes the expected sparsity while ensuring the AMSS of the closed-loop discrete-time system $\Sigma_{cl}$ in Eq.\ref{eqn:sys_cl2} under the sparsification strategies defined in Eq.\ref{compressor} and Eq.\ref{adapcompressor} .
\end{problem}
\end{tcolorbox}

To achieve this goal, we analyze the sequence   $\left(\E(\norm{x(k)}^2),k\geq 0\right)$. The problem can be reduced to finding $K\in \R^{m\times n}$ under which $\lim _{k \rightarrow \infty} \E(\norm{x(k)}^2)~=~0$, for any initial state $x(0)\in \R^n$ while minimizing expected sparsity. This ensures that the system’s trajectories, averaged over many realizations, decay to zero. It indicates that randomness does not cause instabilities that the controller cannot rectify.
\section{Main Results}
In this section, we will primarily focus on the sparsification strategy described in Eq.\ref{compressor}. Then, in the following section, we will discuss Scenario in Eq.\ref{adapcompressor}. The key idea throughout the section is to provide conditions on the parameters of the system $A,B$ and the Bernoulli parameter $p$ under which the dynamics of the expected square norm of the state vector decay geometrically. To this end, we present the following result that will be the basis of our later analysis.
\begin{proposition}{\label{prop}}
    Consider the discrete-time control system $\Sigma$ in Eq.\ref{eqn:sys} under the feedback controller $k\mapsto u(k)=K\mathcal{C}(k)x(k)$, for a gain $K \subseteq \mathbb{R}^{n \times m}$, and the sparsification matrix $k \mapsto \mathcal{C}(k)$ is defined by  $\mathcal{C}(k)= Diag(\frac{c_1(k)}{p}, \ldots,\frac{c_n(k)}{p})$, where $c_i(k)$ are $Ber(p)$ and $p \in (0,1]$. Let $D := A+BK$, $L:=BK$ and $$f(p) := \normm{D^T D + \left(\frac{1-p}{p}\right)\times \text{Diag}(L^{\top}L)}.$$ 
    Then $$\E\left(\norm{x(k+1)}^2\right) \leq  f(p)\E\left(\norm{x(k)}^2\right).$$
Moreover, the following statements hold:
\begin{itemize}
    \item if $f(p)<1$, then the discrete-time control system $\Sigma$ in Eq.\ref{eqn:sys} under the feedback controller $u(k)=K\mathcal{C}(k)x(k)$ is AMSS.
\item if there exists a feedback gain $K$ such that the spectral radius $\rho((A + BK)^\top (A + BK)) < 1$, then there exists $p  \in (0,1)$ such that $f(p) <1$.
\end{itemize}
\end{proposition}

As shown in the previous result, ensuring that $f(p)<1$ makes it possible to guarantee the AMSS of the closed-loop system.  Moreover, a sufficient condition to ensure the existence of such a $p \in (0,1)$, and thus ensuring the existence of a sparsification strategy, is to design a feedback gain $K$ such that all eigenvalues of $(A + BK)^\top (A + BK)$ have norms strictly less than $1$. However, the existence of such a feedback gain $K$ generally requires conditions that are stronger than the classical stabilizability of the pair $(A,B)$~\cite{grune2000homogeneous}. As such, we introduce the following intermediate problem to be addressed in the subsequent discussion.

\begin{remark}
\label{rk}
    As one can see, the map $f(p)$ in Theorem~\ref{sufficient_condition} is decreasing in $p$, indicating that better sparsification (smaller $p$) results in a slower convergence rate of the expected square norm of the state $\E\left(\norm{x(k+1)}^2\right)$, which aligns with the intuition that the controller has access to fewer states at each time step.  
\end{remark}

\begin{tcolorbox}[
  colback=gray!15, colframe=black, boxrule=0.5pt, width=0.48\textwidth,height=0.1\textwidth]
  \begin{problem}
Provide conditions on the pair $(A, B) \in \mathbb{R}^{n \times n} \times \mathbb{R}^{n \times m}$ to guarantee the existence of a matrix $K \in \mathbb{R}^{m \times n}$ such that $\rho((A + BK)^\top (A + BK)) < 1$.
  \end{problem}
\end{tcolorbox}
To answer this question, we introduce the following assumptions for the matrices $A,B\in \R^{n\times n} \times \R^{n\times m}$:
\begin{assumption}
\label{assum:1}
    The matrix $B\in \R^{n\times m}$ is low full rank, i.e $rank(B)=m$.
\end{assumption}
\begin{assumption}
\label{assum:2}
The largest singular value of $(I_n\!-\!B(B^\top B)^{-1}B^\top\!)A$, denoted by $a_n$, is strictly lower than 1.
\end{assumption}
Assumption \ref{assum:2} ensures that the projection matrix $P: =~I_n - B(B^\top B)^{-1} B^\top$  onto the orthogonal complement of the image subspace of $B$ is well-defined. This projection is central as it isolates the part of $A$ that $B$ cannot directly affect. The singular values of $PA$ measure how much effect $A$ has in the directions orthogonal to the image subspace of $B$. The condition $ a_n < 1$  limits the magnitude of $A$’s effect in these orthogonal directions. Intuitively, it says that $A$ does not stretch vectors too much outside the subspace controlled by $B$. This constraint ensures that we can adjust $A$ using $ BK$ to achieve the desired singular values without exceeding certain bounds. It has been established that under these specific assumptions \cite{martin2009singular}, the spectral radius of the matrix $(A + BK)^\top (A + BK)$ is strictly less than one; moreover, these assumptions are both necessary and sufficient for this condition to hold.

Assumptions \ref{assum:1} and \ref{assum:2} are introduced to establish conditions under which it is possible to design a feedback gain matrix  $K$ that ensures the closed-loop system achieves AMSS under randomized sparsification.  To formalize this, we now present the following result, which provides conditions on the existence of such a stabilizing feedback gain.
\begin{proposition}\label{Propo2}
Given $A\in \R^{n\times n}$, $B\in \R^{n\times m}$ and a real number $\gamma$. Consider the following LMI of unknown $K_{\gamma}\in \R^{m\times n}$: 
\begin{equation}{\label{LMI}}
\left[\begin{array}{cc}
\gamma I_n & (A+BK_{\gamma})^T  \\
A+BK_{\gamma} & I_n
\end{array}\right]>0 
\end{equation}
The following holds:
\begin{enumerate}
    \item The  LMI in Eq.\ref{LMI} has solutions if and only if there exists $K_{\gamma}\in ~\R^{m\times n}$ such that $\rho((A + BK_{\gamma})^\top (A + BK_{\gamma})) <\gamma$.
\item If the matrices $A$ and $B$ satisfy Assumptions \ref{assum:1} and \ref{assum:2}, then the LMI in Eq.\ref{LMI} has a solution $K_{\gamma}$ for every $\gamma \in [a_n,1]$.
\end{enumerate}
\end{proposition}
For our application, for each $\gamma \in [a_n,1]$,  we solve the LMI in Eq.\ref{LMI} to determine $K_\gamma$. Interestingly, one should find, for each $K_\gamma$, the smallest $p_\gamma\in (0,1]$ for which $f(p_\gamma)<1$. Consequently, the best strategy is to consider the smallest Bernoulli parameter among all candidates $p_\gamma$ for $\gamma \in [a_n,1]$. To this end, we present the following result.
\begin{theorem}{\label{sufficient_condition}}
    Let $(A, B) \in \mathbb{R}^{n \times n} \times \mathbb{R}^{n \times m}$ satisfying Assumptions \ref{assum:1} and \ref{assum:2}. For $\gamma \in [a_n,1]$, let $K_\gamma$ be the solution of the LMI in Eq.\ref{LMI}. Moreover, let $D_\gamma := A+BK_\gamma$, and $L_\gamma:= BK_\gamma$, and for all $i\in \{1,\ldots,n\}$, let $ s_i :=\sum_{k=1}^{n} l_{k,i}^2$, and $s_{\text{max}}~=~\max\left(s_1,\ldots,s_n\right)$. Furthermore, let $\alpha_\gamma:=~\frac{1-\normm{D_\gamma}^2}{s_{\text{max}}}$. If the parameter $p$ satisfies:
    \begin{equation}{\label{Pkgamma}}
        p> p_{K_\gamma} :=\frac{1}{1+\alpha_\gamma},
    \end{equation} 
    then, the system defined by Eq.\ref{eqn:sys_cl2} under sparsification defined by the sparsification matrix $k\mapsto \C(k)$ in Eq.\ref{compressor} is asymptotically mean-square stable.
\end{theorem}

We propose an algorithm based on Theorem \ref{sufficient_condition}. The inputs to Algorithm \ref{algo1} are $ A$ and $B$ matrices of the dynamics satisfying Assumptions \ref{assum:1} and \ref{assum:2}. The algorithm computes $a_n$, the largest singular value of the matrix $\left(I_n - B(B^\top B)^{-1} B^\top \right)A$, discretizes the interval $[a_n,1]$ with step size $\delta$ to a list of discrete points $\mathcal{L}$. Then, it loops over the list $\mathcal{L}$ and solves the LMI in Eq.\ref{LMI}. As such, it obtains a list of control gain matrices $K_\gamma$ for $\gamma\in \mathcal{L}$, which are required to compute $p_{K_\gamma}$ using the formula in Eq.\ref{Pkgamma} in Theorem \ref{sufficient_condition}. It returns, as the last step, $p^\star$, the smallest sparsification parameter, and its corresponding gain matrix $K.$
{
\begin{remark}
{\label{ass}}
 Building on Theorem 3.11 in \cite{sun2011stability}, one can show that AMSS ensures the almost sure stability of the discrete-time control system $\Sigma$ in Eq.\ref{eqn:sys_cl2}  under the random sparsification mechanism defined in Eq.\ref{compressor}.
\end{remark}}
\begin{algorithm}[tb]
\caption{Compute the control gain matrix $K$ and Bernoulli parameter $p^\star$}
\label{algo1}
\textbf{Input}: Matrices $A \in \mathbb{R}^{n \times n}$, and $B \in \mathbb{R}^{n \times m}$ that satisfy Assumptions \ref{assum:1} and \ref{assum:2}.\\
\textbf{Output}: Bernoulli parameter $p\in (0,1]$ and control gain matrix $K\in \R^{m\times n}$.
\begin{algorithmic}[1] 
\STATE \textbf{Compute} $a_n\in (0, 1]$, the greatest singular value of the matrix $\left(I_n - B(B^\top B)^{-1} B^\top \right)A$.
\STATE \textbf{Discretize} the interval $[a_n,1]$ with step-size $\delta$ into a list $\mathcal{L}$.
\FOR{various values of $\gamma \in \mathcal{L}$}
\STATE \textbf{Compute} the corresponding gain $K_\gamma$ by solving the LMI in Eq.\ref{LMI}.
\STATE \textbf{Compute} $p_{K_\gamma}$ from Eq.\ref{Pkgamma} in Theorem~\ref{sufficient_condition}.
\ENDFOR
\STATE \textbf{Return}  $p^{\star}=\text{Min}_{\gamma \in \mathcal{L}}\left(p_{K_\gamma} \right)$ and the corresponding $K$.
\end{algorithmic}
\end{algorithm}

\subsection{Adaptive sparsification}

While the previous section addressed stability under a uniform sparsification strategy, real-world systems often face heterogeneous sensing costs. This motivates a more practical, component-wise approach, where sparsity levels are tailored to each state component. We therefore analyze the adaptive sparsification strategy defined in Eq.\ref{adapcompressor}, and jointly design the gain matrix 
$K$ and parameters 
$p_1,\ldots,p_n$ to ensure AMSS while minimizing expected sparsity.
Accordingly, we extend the results of Proposition \ref{Propo2} to this more general setting, where each state component is accessed with its own probability. This leads to the following result.


\begin{proposition}{\label{propegene}}
Consider the discrete-time control system $\Sigma$ in Eq.\ref{eqn:sys_cl2} under the feedback controller given by $u(k)=K\mathcal{C}(k)x(k)$ for $K \subseteq \mathbb{R}^{n \times m}$ and $\mathcal{C}(k)$ defined in Eq.\ref{adapcompressor}. Let $D := A+BK$,$L:=BK$, and 
{\small
$$g(p_1,\ldots,p_n)=  \normm{D^{\top}D+\!Diag\left(-s_1+\!\frac{s_1}{p_1},\ldots,-s_n+\!\frac{s_n}{p_n}\right)}.\!$$} The following holds for all $k \in \mathbb{N}$, 
\begin{equation*}
    \E(\norm{x(k+1)}^2) \leq g(p_1,\ldots,p_n) \E(\norm{x(k)}^2)
\end{equation*}
Moreover, if $g(p_1,\ldots,p_n)<1$, then the system $\Sigma$ in Eq.\ref{eqn:sys} under the feedback controller $u(k)=K\mathcal{C}(k)x(k)$, where the compressor $\mathcal{C}(k)$ defined in Eq.\ref{adapcompressor}, is AMSS.
\end{proposition}
This result illustrates that it is sufficient to design $K$ and Bernoulli parameters $p_1,\ldots,p_n$ to have $g(p_1,\ldots,p_n)<1$ and by that guarantee the AMSS of the system. To achieve this goal, we assume that the dynamics of the system $A$ and $B$ satisfy Assumptions \ref{assum:1} and \ref{assum:2}. Moreover, we propose Algorithm \ref{algo2} based on the following result:
\begin{theorem}{\label{mainadaptive}}
 Let $(A, B) \in \mathbb{R}^{n \times n} \times \mathbb{R}^{n \times m}$ satisfying assumptions \ref{assum:1} and \ref{assum:2}. For $\gamma \in [a_n,1]$, let $K_\gamma$ be the solution of LMI in Eq.\ref{LMI}. Moreover, let $D_\gamma := A+BK_\gamma$, and $L_\gamma:= BK_\gamma$, and for all $i\in \{1,\ldots,n\}$, let $ s_i :=\sum_{k=1}^{n} l_{k,i}^2$. If for all $i\in \{1,\ldots,n\}$, $$p_i> p_{i,K_\gamma} : = \frac{1}{1+\frac{1-\normm{D_\gamma}^2}{s_i}},$$ 
 then, the system defined by Eq.\ref{eqn:sys_cl2} under sparsification strategy $k \mapsto \mathcal{C}(k)$ in Eq.\ref{adapcompressor} is asymptotically mean-square stable.
\end{theorem}
\begin{algorithm}[tb]
\caption{Compute the Gain Matrix K and Bernoulli parameter [$p^{\star}_1,\ldots,p^{\star}_n$]}
\label{algo2}
\textbf{Input}: Matrices $A \in \mathbb{R}^{n \times n}$, and $B \in \mathbb{R}^{n \times m}$ that satisfy Assumptions \ref{assum:1} and \ref{assum:2}, and weights $w_1,\ldots,w_n.$\\
\textbf{Output}: sparsification parameter $\left[p_1,\ldots,p_n\right]$, control gain matrix $K$.
\begin{algorithmic}[1] 
\STATE \textbf{Compute} $a_n\in (0, 1]$, the greatest singular value of the matrix $\left(I_n - B(B^\top B)^{-1} B^\top \right)A$.
\STATE \textbf{Discretize} the interval $[a_n,1]$ with step-size into a list $\mathcal{L}$.
\FOR{various values of $\gamma \in \mathcal{L}$}
\STATE \textbf{Compute} the corresponding gain $K_\gamma$ by solving the LMI in Eq.\ref{LMI}.
\STATE \textbf{Compute} $p_{1,K_\gamma},\ldots,p_{n,K_\gamma} $ using the formula in Theorem~\ref{mainadaptive}.
\STATE \textbf{Compute} the expected sparsity $ES_{\gamma}:=~ \sum_{i=1}^{n} w_i p_{i,K_\gamma}$
\ENDFOR
\STATE \textbf{Return}  $\left[p^{\star}_1,\ldots,p^{\star}_n\right]= \text{Min}_{\gamma \in \mathcal{L}}\left(ES_{\gamma}\right)$, the corresponding $K$.
\end{algorithmic}
\end{algorithm}
Theorem \ref{mainadaptive} provides a sufficient condition to ensure that the closed-loop system operates under AMSS. It quantifies the influence of each individual state component on the control input via the term 
 $s_i$, which represents the squared norm of the i-th column of $BK_\gamma$. Specifically, a small $s_i$ allows for a smaller $p_i$, meaning that the controller can drop the corresponding state component $x_i$ more frequently. In contrast, a significant $s_i$ leads to a large $p_i$, indicating that the corresponding state component significantly affects control performance and should be sampled more often. Moreover, $\normm{D_\gamma}^2$ captures how close the system is to stability. A lower norm of $D_\gamma=A+K_\gamma B$ indicates a more stable system and enables better sparsification without compromising AMSS.

Based on Theorem~\ref{mainadaptive}, Algorithm~\ref{algo2} computes the sparsification vector $[p_{1, K_\gamma},\ldots,p_{n, K_\gamma}]$ that minimizes expected sparsity $ES$ and returns the corresponding gain matrix $K$.
\begin{remark}
[Link to observability]
We explain how our framework naturally extends to the case where the controller receives a sparsified version of the reconstructed state. 
Consider the discrete-time linear system
\begin{equation}
\label{eq:plant_masked}
\Sigma_o:\;
\begin{cases}
x(k+1) = A x(k) + B u(k),\\[2mm]
z(k) = C x(k),
\end{cases}
\end{equation}
where $x(k)\in\mathbb{R}^{n}$ is the state, $u(k)\in\mathbb{R}^{m}$ is the control input, and $z(k)\in\mathbb{R}^{r}$ is the measured output. The matrices $A,B,C$ have compatible dimensions.
Assuming $(A,B)$ is stabilizable and $(A,C)$ is observable/detectable, the separation principle \cite{aastrom2021feedback} ensures a Luenberger observer can reconstruct $x(k)$ from $z(k)$ while preserving closed-loop stability. Our randomized sparsification can then be applied to the estimated state, maintaining the theoretical guarantees of full-state feedback.
\end{remark}
\section{Numerical Results}
We validate our analysis on benchmark discrete-time control systems. To approximate the expected state trajectory $\mathbb{E}[x(k)| x(0)]$, we use Monte Carlo simulations: multiple trajectories are generated from a common initial state $x(0)$, each subject to independent realizations of the randomized sparsification strategies in Eq.\ref{compressor} and Eq.\ref{adapcompressor}. The expected trajectory is estimated as the empirical mean over $N=100$ runs, with $x(0)$ sampled from a zero-mean Gaussian distribution with variance $\sigma =100$. Additional experiments are provided in Appendix B. All simulations were conducted in Python using the CVX toolbox \cite{cvx} with MOSEK as the solver \cite{aps2022mosek}.

 \begin{figure}[t]
    \centering
\includegraphics[width=\linewidth]{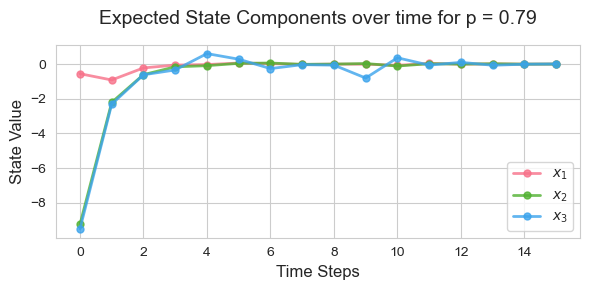}
    \caption{Time evolution of the expected state components. The figure depicts stability for the case where $p = p^{\star} = 0.79$}
    \label{fig:example1_1}
\end{figure}
\begin{figure}
\includegraphics[width=\linewidth]{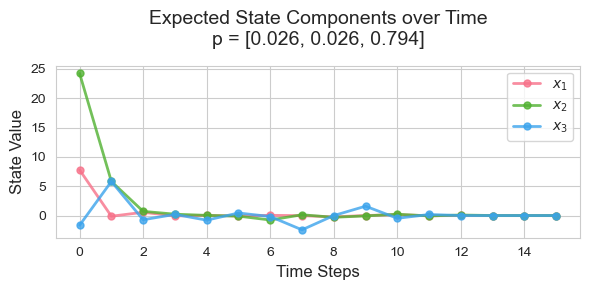}
    \caption{Time evolution of the expected state components. The Figure Shows the stability of every component of the expected state under the adaptive sparsification strategy.}
    \label{fig:example1_3}
\end{figure}

\begin{figure*}[t]
\centering
    \begin{subfigure}[t]{0.48\textwidth}   \includegraphics[width=\linewidth,height=4cm]{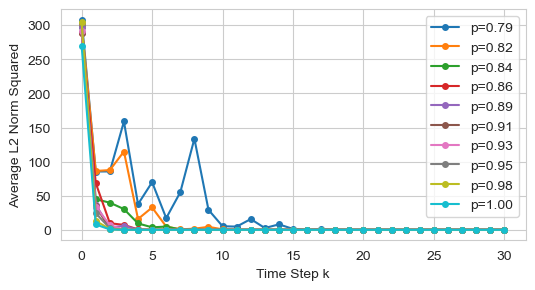}
    \caption{ The system's AMSS is satisfied when the sparsification parameter meets or exceeds  $p^\star$.}
    \label{fig:example1_4}  
    \end{subfigure}
    \hfill
    \begin{subfigure}[t]{0.48\textwidth}
        \centering       \includegraphics[width=\linewidth,height=4.3cm]{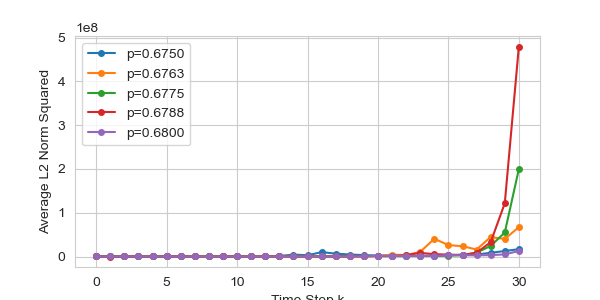}
    \caption{ The system's AMSS is not fulfilled for values of the sparsification parameter strictly lower than $p^\star$.}
    \label{fig:example1_AMSS}  
    \end{subfigure}
    \caption{Variation of $\E_{x(0),\C}\left(\norm{x(k)}^2\right)$ displayed for different values of the sparsification parameter.}
\end{figure*}
\subsection{Grid-Forming Converter}
We consider a grid-forming converter from \cite{chen2023power}. We use our approach on a slightly modified dynamics where the matrices $A$ and $B$ are given by
\[
A = \begin{bmatrix}
0 & 0 & 0.1017\\
0 & 0 & 0.025 \\
0 & 0 & 2
\end{bmatrix}, \qquad B = \begin{bmatrix}
1 & 0.005 \\
0 & 1.5095 \\
314.1593 & 0
\end{bmatrix}.
\]
Algorithm \ref{algo1} allows to compute the Bernoulli parameter used for the sparsification strategy given by  $p^\star=0.79$ along with the control gain matrix $K$.  Fig.\ref{fig:example1_1} illustrates that all three components of the expected state converge to $0$,  which is consistent with the theoretical results in Lemma \ref{expected_stability} and Theorem \ref{sufficient_condition}. Fig.\ref{fig:example1_3} shows the variation of expected state components with the sparsification vector $p^\star = [ 0.026,0.026,0.794]$ resulting from our Algorithm \ref{algo2}. Finally, Fig.~\ref{fig:example1_4} illustrates the evolution of the expected squared two-norm of the state based on a Monte Carlo simulation, showing that with any sparsification strategy with  ($p \geq p^\star = 0.79$), the system achieves AMSS. Moreover, one can also see the effect of sparsification on the convergence rate; higher sparsification implies slower convergence rates, which follows the analytical insights provided in Remark \ref{rk}. Fig.\ref{fig:example1_AMSS} depicts that, for values of $p$ strictly lower than $p^\star$, the AMSS of the system is not fulfilled. 

\subsection{Large-scale  power system}
We consider the model of a large-scale power system from \cite{alonso2022distributed}. It describes the dynamics of a fully connected network of \(n\) nodes. The discrete-time state equations for the i-th node are given by:
\begin{align*}
\theta_i(k\!+\!1)\!= & \theta_i(k) + \Delta k \cdot \omega_i(k) + b_{1,i} u_{1,i}, \\
\omega_i(k\!+\!1)\!= &\!\!\sum_{\substack{j=1 \\ j \neq i}}^n\!\! \frac{k_{ij}}{m_i}\! \Delta k \!\cdot\! (\!\theta_j(\!k)\! - \!\theta_i(\!k)\!)\! 
+\!\!\alpha_i\!\! \sum_{j=1}^n \!\!\omega_j(k)\! +\! b_{2,i}\! u_{2,i},
\end{align*}
for \(i = 1, 2, \ldots, n\), where \(\theta_i(k)\) is the phase angle of node \(i\), \(\omega_i(k)\) is the frequency deviation of node \(i\), \(m_i\) is the inertia parameter of node \(i\), $\alpha_i=1-\!\frac{d_i}{m_i}\! \Delta k$ where $d_i$ is the damping parameter of node \(i\), $k_{ij}$ is the coupling strength between nodes \(i\) and \(j\), with \(k_{ii} = 0\), and \(\Delta k\) the sampling period. The global state vector is given by $x(k) := \left[\theta_1(k),\omega_1(k),\ldots,\theta_n(k),\omega_n(k)\right]^{\top}$. We write the dynamics of $x(k)$ in the form of Eq.\ref{eqn:sys}, where the control input matrix $B$ is a $2n \times 1$ input vector given by $B = \left[b_{1,1},b_{2,1},\ldots,b_{1,n},b_{2,n}\right]^{\top}$, and the matrix $A$ is a $2n \times 2n$  matrix constructed as follows: (1) For each row block $i = 1, \dots, n$, concatenate the submatrices $A_{ij}$ (for $j = 1, \dots, n$) horizontally to form a $2 \times 2n$ block row, (2) concatenate these $n$ block rows vertically to form the full $2n \times 2n$ matrix $A$ with $A_{ii} = \begin{bmatrix}
        1 & \Delta k \\
        -\frac{k_i}{m_i} \Delta k & 1 - \!\frac{d_i}{m_i} \Delta k
    \end{bmatrix}$ and $ \!A_{ij} = \begin{bmatrix}
        0 & 0 \\
        \frac{k_{ij}}{m_i} \Delta k & 1 -\! \frac{d_i}{m_i} \Delta k
    \end{bmatrix}$.
For numerical simulation, we consider $n=1000$ nodes and use \(\Delta k = 0.2\), the numerical values of the masses $m_i$ (and damping parameters \(d_i\)), $i=1,\ldots,1000$, are generated from a uniform distribution over the interval \([0.5, 2.0)\) ( \([0.5, 1.0)\), respectively). 
\begin{figure}[t]
    \centering
\includegraphics[width=\linewidth,height=4cm]{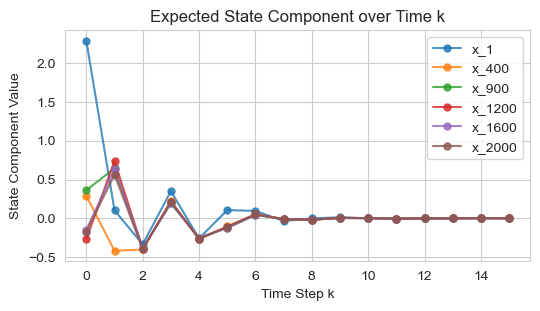}
    \caption{Time evolution of the expected state components. It depicts stability for the case where $p = p^{\star} = 0.0026$.}
    \label{fig:example3_1}
\end{figure}
\begin{figure}
    \centering    \includegraphics[width=\linewidth,height=4cm]{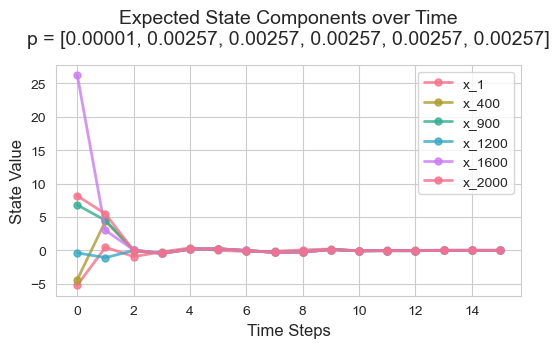}
    \caption{Stability of components $\{1,400,900,1200,1600,2000\}$ of the expected state under the adaptive sparsification strategy.}
    \label{fig:example3_3}
\end{figure}
\begin{figure}[H]
    \centering
\includegraphics[width=\linewidth,height=3.6cm]{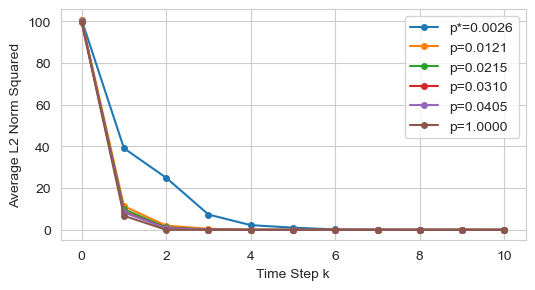}
\caption{$\E_{x(0),\C}\left(\norm{x(k)}^2\right)$ for different values of the sparsification parameter $p\geq p^\star$, with AMSS satisfied.}
  \label{fig:example3_4}
\end{figure}
Using Algorithm \ref{algo1}, we determine the Bernoulli parameter  $(p^\star=0.0026)$ and the associated control gain matrix  $K$. Fig.\ref{fig:example3_1} displays the evolution of the expected state components, which converge to $0$, which is consistent with the result in Lemma \ref{expected_stability}. Fig.\ref{fig:example3_3} shows the variation of the expected state components with the sparsification vector $p^\star = [p_1^\star,\ldots,p_n^\star]$ resulting from Algorithm \ref{algo2}. Finally, we compare the expected square norm of the state vector for different values of parameter $p$ from $p^\star$ to $p=1$ (classical stabilization based on full state measurement without sparsification). The results shown in Fig.\ref{fig:example3_4} align with the theoretical result in Theorem \ref{sufficient_condition}. {{Simulations show that deterministic approaches require at least $37$ state measurements for full state reconstruction, whereas our randomized method achieves AMSS with only $6$ measurements. Moreover, finding such minimal measurement sets is NP-hard \cite{olshevsky2015minimum}, as it involves combinatorial observability verification.}}

\section{Conclusion}
In this work, we studied the AMSS of linear discrete-time systems under a randomized feedback strategy. We showed, for the first time, that AMSS can be ensured using only a randomly selected subset of state measurements. We proposed a procedure to jointly design the feedback gain and sparsification parameters that balance stability and sensing efficiency. We extended our analysis to adaptive sparsification, where access probabilities are tailored to state components. Our experimental results showed that AMSS can be achieved using only $0.3\%$ of the state components in some settings. Future work will extend the proposed framework in several directions. First, we will assess the tightness of our inequality-based derivations and the conservativeness of $p^\star$. Second, we will extend our analysis to the case of noisy systems. Finally,  we will generalize the proposed approach from stabilization to more complex control objectives, such as those described using temporal logics \cite{saoud2024temporal,han2020linear}.
\section{Appendix A}
This section presents the proofs of our theoretical results, accompanied by a brief discussion on the complexity of algorithms \ref{algo1} and \ref{algo2}.
\subsection{Proof of Lemma \ref{expected_stability}}
 We have $x(k+1)=(A+B K \mathcal{C}(k)) x(k)$ for all $k \in \mathbb{N}$. By applying the conditional expectation, one gets:
 $$\mathbb{E}(x(k+1)\mid x(k))=(A+BK\mathbb{E}(\mathcal{C}(k))) x(k)$$ 
 Reapplying the expectation yields 
 \begin{align*}
    \mathbb{E}(x(k+1))&=(A+B K \mathbb{E}(\mathcal{C}(k))) \mathbb{E}(x(k))\\
     &=(A+B K) \mathbb{E}(x(k))
 \end{align*}
 The last equality follows from the fact that for each time step $k$, we have $\mathbb{E}(\mathcal{C}(k))=I_n$, which completes the proof.
\subsection{Proof of proposition \ref{prop}}
      For each time step $k$, the system Eq.\ref{eqn:sys_cl2} can be written as $x(k+1) = M(k)x(k)$ where $M(k) :=A+~BK\C(k) $. By applying the expectation, we get:
 \begin{align*}
     \E\left(\norm{x(k+1)}^2\mid x(k) \right) &= x(k)^{\top}\E\left(M(k)^{\top}M(k)\right)x(k)\\
    &\leq \!\norm{\!x(k)}\! \norm{\E\!\left(M\!(k)^{\top}\! M\!(k)\!\right)\! x(k)} \\
     &\leq \norm{x(k)}^2 \normm{\E\left(M(k)^{\top}M(k)\right)}
 \end{align*}
 where the first inequality is the Cauchy-Schwarz inequality, and the last one follows from the submultiplicativity of the $\norm{.}$. By retaking the expectation, we get:
 \begin{equation}
 \label{eqn:2}
     \E(\norm{x(k+1)}^2) \leq  \normm{\E\left(M(k)^{\top}M(k)\right)}\E(\norm{x(k)}^2).
 \end{equation}
 Let us now analyze the term $\E\left(M(k)^{\top}M(k)\right)$.
 \begin{align*}{\label{theterm}}
    \E\left(M(k)^{\top}M(k)\right)&=\E\left( \left(A^{\top} + \C(k)^{\top}L^{\top}\right)\left(A+L\C(k)\right) \right) \\
    & = A^{\top} A + A^{\top} L+L^{\top}A+\\
     &\E\left(\C(k)^{\top}L^{\top} L\C(k)\right),
 \end{align*}
 where the last equality follows from the fact that $$ \mathbb{E}(\mathcal{C}(k))=I_n.$$
 Let $i,j \in \{1,\ldots,n\}$ and time step $k$. We have:  $(\C(k)^{\top}L^{\top} L\C(k))_{i,j} = \sum_{r=1}^{n} l_{r,i}l_{r,j}\frac{ c_i(k)c_j(k)}{p^2}$. We distinguish two cases: 
 \begin{itemize}
    \item If $i\ne j$, then  for all $k \in \mathbb{N}$ one gets $\E(c_i(k)c_j(k))=\E(c_i(k))\E(c_j(k)) = p^2$, because $c_i(k)$ and $c_j(k)$ are independent. Which implies that 
     $\mathbb{E}\left((\C(k)^{\top}L^{\top} L\C(k))_{i,j}\right) = (L^{\top} L)_{i,j} $
     \item  If $i=j$, then for all $k\in \mathbb{N}$,  one has $\E(c_i^2(k))=p.$ Thus, 
 \begin{align*}
 \mathbb{E}((\C(k)^{\top}L^{\top} L\C(k)))_{i,i} &=  \frac{1}{p} \sum_{r=1}^{n} l_{r,i}l_{r,i}= \frac{1}{p}(L^\top L)_{i,i}\\
 &= (L^\top L)_{i,i} + \frac{1-p}{p} (L^\top L)_{i,i}.  
 \end{align*}
\end{itemize}
 We conclude that 
 $$\E\left(\C(k)^{\top}L^{\top} L\C(k)\right) = L^{\top} L + \frac{1-p}{p}Diag\left(L^{\top} L\right).$$
 Therefore, one gets:
 $$\E\left(M(k)^{\top}M(k)\right) = D^\top D + \frac{1-p}{p}Diag\left(L^{\top} L\right),$$
 which in turn implies from Eq.\ref{eqn:2} that:
 \begin{equation}{\label{geometric}}
     \E\left(\norm{x(k+1)}^2\right) \leq  f(p)\E\left(\norm{x(k)}^2\right).
 \end{equation}
Clearly, if $f(p) < 1$, then $\left(\E(\norm{x(k)}^2),k\geq 0\right)$ exhibits a geometric decay, which implies that the discrete-time control system $\Sigma$ in Eq.\ref{eqn:sys} under the feedback controller $u(k)=K\mathcal{C}(k)x(k)$ is AMSS.
To show the last item we consider the map $p\mapsto g(p)$ defined by:
 $$g(p)=\normm{D}^2+ \left(\frac{1}{p}-1\right)\times \normm{\text{Diag}(L^{\top}L)}.$$
Using the triangular inequality we have for all $p \in (0,1]$, 
$f(p)  \leq g(p)$. Since there exits a feedback gain $K$ such that the spectral radius $\rho((A + BK)^\top (A + BK)) < 1$ it follows that $g(1)=\normm{D}^2<1$. Moreover, using the fact that the map $p \mapsto g(p)$ is continuous on $p$, we have the existence of $p \in (0,1]$ such that $f(p) \leq g(p) <1$.
\subsection{Proof of proposition \ref{Propo2}}
 The first result is a direct consequence of Schur complement, see \cite{zhang2006schur}.  For the second point, it follows from 
 Theorem 2.7 in \cite{martin2009singular}, which states that it is possible to assign singular values of $A+BK_{\gamma}$ that are lower than $1$ as long as the largest singular value of $(I_n-B(B^\top B)^{-1}B^\top)A$ is strictly lower than $1$, which completes the proof.
\subsection{Proof of Theorem \ref{sufficient_condition}}
 Let $\gamma\in [a_n,1]$ and $p > \frac{1}{1+\alpha_\gamma}$ where $\alpha_\gamma = \frac{1-\normm{D_\gamma}^2}{s_{\text{max}}}$. Therefore, one gets
 \begin{equation}
\label{eqn:pr2}
     \frac{1-p}{p} <\frac{1-\normm{D_\gamma}^2}{s_{\text{max}}}.
 \end{equation}
 Using the definition of the spectral norm of a matrix, it follows that
{\small \begin{align*}
  &\normm{D_{\gamma}^T D_{\gamma} + \left(\frac{1-p}{p}\right)\times \text{Diag}(L_{\gamma}^{\top}L_{\gamma})}\\&=\text{sup}_{\norm{x}=1} \norm{{(D_{\gamma}^T D_{\gamma} + \left(\frac{1-p}{p}\right)\times \text{Diag}(L_{\gamma}^{\top}L_{\gamma}))x}} \\
  &\leq \text{sup}_{\norm{x}=1} \norm{{D_{\gamma}^T D_{\gamma}x}}+\\
  &\text{sup}_{\norm{x}=1} \norm{{ \left(\frac{1-p}{p}\right)\times \text{Diag}(L_{\gamma}^{\top}L_{\gamma})x}} \\
 & < \normm{D_{\gamma}}^2+\text{sup}_{\norm{x}=1} \norm{{ \left(\frac{1-\normm{D_\gamma}^2}{s_{\text{max}}}\right)\times \text{Diag}(L_{\gamma}^{\top}L_{\gamma})x}}\\
   &< \normm{D_{\gamma}}^2+\\
   &  \left(1-\normm{D_\gamma}^2\right)\times \text{sup}_{\norm{x}=1} {\norm{\text{Diag}\left(\frac{s_1}{s_{\text{max}}},\ldots,\frac{s_n}{s_{\text{max}}})\right)x}}\\
   &< \norm{D_{\gamma}}^2+{ \left(1-\normm{D_\gamma}^2\right)\normm{ \text{Diag}\left(\frac{s_1}{s_{\text{max}}},\ldots,\frac{s_n}{s_{\text{max}}})\right)}}
   \leq 1
 \end{align*}}
where the first inequality comes from the use of the triangular inequality, the second inequality comes from (\ref{eqn:pr2}), the third inequality uses the fact that $Diag(L_{\gamma}^{\top}L_{\gamma})=Diag(s_1,s_2,\ldots,s_n)$ and the foorth inequality follows from the fact that $s_{\max}=\max(s_1,s_2,\ldots,s_n)$. Hence one gets $\normm{D_{\gamma}^T D_{\gamma} + \left(\frac{1-p}{p}\right)\times \text{Diag}(L_{\gamma}^{\top}L_{\gamma})} <1$, which in turn implies that the system defined by Eq.\ref{eqn:sys_cl2} under sparsification defined by the sparsification matrix $k\mapsto \C(k)$ in Eq.\ref{compressor} is asymptotically mean-square stable.
\subsection{Proof of proposition \ref{propegene}}
      For each time step $k$, the system Eq.\ref{eqn:sys_cl2} can be written as $x(k+1) = M(k)x(k)$ where $M(k) :=A+~BK\C(k) $. By applying the expectation, we get:
 \begin{align*}
     \E\left(\norm{x(k+1)}^2\mid x(k) \right) &= x(k)^{\top}\E\left(M(k)^{\top}M(k)\right)x(k)\\
    &\leq \!\norm{\!x(k)}\! \norm{\E\!\left(M\!(k)^{\top}\! M\!(k)\!\right)\! x(k)} \\
    & \quad\\
     &\leq \norm{x(k)}^2 \normm{\E\left(M(k)^{\top}M(k)\right)}
 \end{align*}
 where the first inequality is the Cauchy-Schwarz inequality, and the last one follows from the submultiplicativity of the $\norm{.}$. By retaking the expectation, we get:
 \begin{equation}
 \label{eqn:2}
     \E(\norm{x(k+1)}^2) \leq  \normm{\E\left(M(k)^{\top}M(k)\right)}\E(\norm{x(k)}^2).
 \end{equation}
 Let us now analyze the term $\E\left(M(k)^{\top}M(k)\right)$.
 \begin{align*}{\label{theterm}}
    \E\left(M(k)^{\top}M(k)\right)&=\E\left( \left(A^{\top} + \C(k)^{\top}L^{\top}\right)\left(A+L\C(k)\right) \right) \\
    & = A^{\top} A + A^{\top} L+L^{\top}A\\
     &+\E\left(\C(k)^{\top}L^{\top} L\C(k)\right),
 \end{align*}
 where the last equality follows from the fact that $$ \mathbb{E}(\mathcal{C}(k))=I_n.$$
 Let $i,j \in \{1,\ldots,n\}$ and time step $k$. We have:  $(\C(k)^{\top}L^{\top} L\C(k))_{i,j} = \sum_{r=1}^{n} l_{r,i}l_{r,j}\frac{ c_i(k)c_j(k)}{p_i\times p_j}$. We distinguish two cases: 
 \begin{itemize}
    \item If $i\ne j$, then  for all $k \in \mathbb{N}$ one gets $\E(c_i(k)c_j(k))=\E(c_i(k))\E(c_j(k)) = p_i \times p_j$, because $c_i(k)$ and $c_j(k)$ are independent. Which implies that 
     $\mathbb{E}\left((\C(k)^{\top}L^{\top} L\C(k))_{i,j}\right) = (L^{\top} L)_{i,j} $
     \item  If $i=j$, then for all $k\in \mathbb{N}$,  one has $\E(c_i^2(k))=p_i.$ Thus, 
 \begin{align*}
 \mathbb{E}((\C(k)^{\top}L^{\top} L\C(k)))_{i,i} &=  \frac{1}{p} \sum_{r=1}^{n} l_{r,i}l_{r,i}= \frac{1}{p}(L^\top L)_{i,i}\\
 &= (L^\top L)_{i,i} + \frac{1-p_i}{p_i} (L^\top L)_{i,i}  
 \end{align*}
\end{itemize}
 We conclude that 
{\footnotesize  \begin{align*}
     \E\left(\C(k)^{\top}L^{\top} L\C(k)\right) &= L^{\top} L+ \\
     &Diag\!\left(\!-s_1\!+\!\frac{1\!-\!p_1}{p_1}, \ldots,\!-\!s_n\!+\!\frac{1\!-\!p_n}{p_n}\right)
 \end{align*}}
 Therefore, one gets:
 $$\E\left(\!M(k)^{\top}\!M(k)\!\right)\! =\! D^\top \!D \!+\! Diag\!\left(\!\!-s_1\!+\!\frac{s_1}{p_1},\ldots,\!-\!s_n\!\!+\!\frac{s_n}{p_n}\!\!\right) $$
 which in turn implies from Eq.\ref{eqn:2} that:
$$\E\left(\norm{x(k+1)}^2\right) \leq  g(p_1,\ldots,p_n)\E\left(\norm{x(k)}^2\right).$$
\subsection{Proof of Theorem \ref{mainadaptive}}
 Let $\gamma \in [a_n,1]$ and suppose that for all $i\in \{1,\ldots,n\}$,  $p_i>\frac{1}{1+\frac{1-\normm{D_\gamma}^2}{s_i}}$. Therefore for all $i\in \{1,\ldots,n\}$, we have,  $$-s_i +\frac{s_i}{p_i} +\normm{D_\gamma}^2< 1.$$ Let $y = [y_1,\ldots,y_n]^\top\in \R^{n}$, we have:
    \begin{align*}
        &y^\top (Diag(-s_1+\frac{s_1}{p_1},\ldots, -s_n+\frac{s_n}{p_n}) + D_\gamma^\top D_\gamma)y \\
        & = y^\top (Diag(-s_1+\frac{s_1}{p_1},\ldots, -s_n+\frac{s_n}{p_n})y + y^\top D_\gamma^\top D_\gamma y \\
        & \leq y^\top (Diag(-s_1+\frac{s_1}{p_1},\ldots, -s_n+\frac{s_n}{p_n})y +  \normm{D_\gamma}^2\norm{y}^2 \\
        &\leq \sum_{i=1} ^{n} (-s_i+\frac{s_i}{p_i})y_i^2+ \sum_{i=1} ^{n}  \normm{D_\gamma}^2 y_i^2 \\
        & \leq \sum_{i=1} ^{n} (-s_i+\frac{s_i}{p_i} + \normm{D_\gamma}^2) y_i^2 \\
        & < \norm{y}^2
    \end{align*}
    Thus, $$D_\gamma^{\top}D_\gamma + Diag(-s_1+\frac{s_1}{p_1},\ldots, -s_n+\frac{s_n}{p_n}) < I_n, $$
    
     from which follows that 
     $$\normm{D_\gamma^{\top}D_\gamma+Diag\left(-s_1+\frac{s_1}{p_1},\ldots, -s_n+\frac{s_n}{p_n}\right)} <1.$$
     Therefore $g(p_1,\ldots,p_n)<1$, which in turn implies that the system defined by Eq.\ref{eqn:sys_cl2} under sparsification defined by the sparsification matrix $k\mapsto \C(k)$ in Eq.\ref{adapcompressor} is asymptotically mean-square stable.
\subsection{On the complexity of the Algorithms}
Given the matrices $A\in \mathbb{R}^{n\times n}$ and $B \in \mathbb{R}^{n \times m}$, the performance of our algorithms is significantly affected by two key steps, outlined as follows:
\begin{enumerate}
    \item \textbf{Compute \( a_n \), the greatest singular value of \ $\left(I_n - B(B^\top B)^{-1} B^\top \right)A$:} The computation of the product 
$\left(I_n - B(B^\top B)^{-1} B^\top \right)A$ is dominated by the matrix inversion $(B^\top B)^{-1}$,  which has a complexity of $\mathcal{O}(n^3)$. Furthermore, computing the largest singular value of the resulting matrix also has a worst-case complexity of $\mathcal{O}(n^3)$ \cite{golub2013matrix}. Therefore, the overall complexity of this step is $\mathcal{O}(n^3)$.
    \item \textbf{Solving an LMI via semidefinite programming (SDP):} 
    Solving the LMI has a worst-case complexity of \( \mathcal{O}(n^4) \) when using standard interior-point solvers \cite{nesterov1994interior}.
\end{enumerate}
The overall complexity is thus \( \mathcal{O}(n^4) \), dominated by the LMI solution. It is worth noting that in control theory, stabilization typically involves solving LMIs using convex optimization solvers. Additionally, the optimization of the sparsification strategy is performed offline, once the outputs of the algorithm are computed (which corresponds to the gain matrix $K$ and the Bernoulli parameter $p^*$), the system operates efficiently online using the resulting sparsified vectors.

\begin{figure}[t]
    \centering    \includegraphics[width=\linewidth,height=4cm]{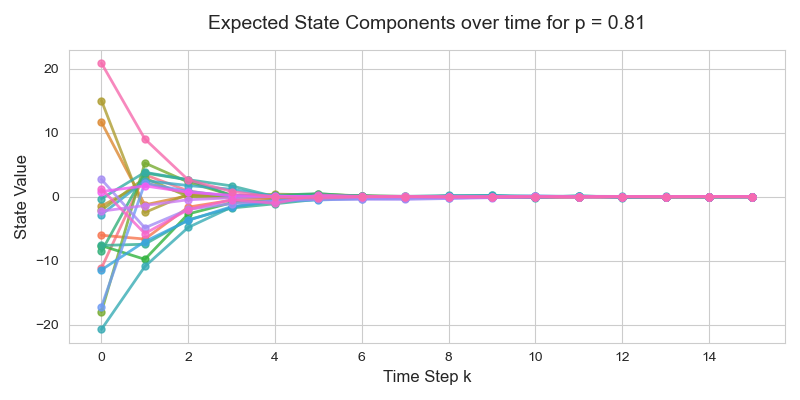}
    \caption{Time evolution of the expected state components. It depicts stability for the case where $p = p^{\star} = 0.81$.}
    \label{fig:example2_1_1}
\end{figure}
\begin{figure}[t]
    \centering    \includegraphics[width=\linewidth,height= 4cm]{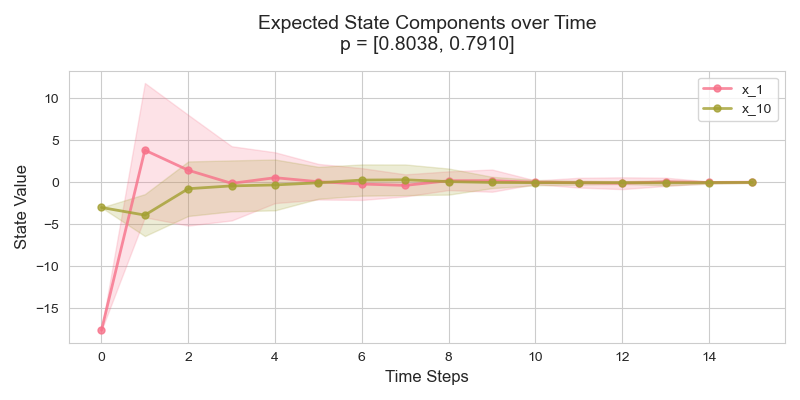}
    \caption{The stability of components $1$ and $10$ of the expected state under the adaptive sparsification strategy with parameters $0.8038$ and $0.791$ successively. Solid lines represent the mean values, and the $±$ standard deviations are shown as shaded contours.}
    \label{fig:example2_3_A}
\end{figure}
\begin{figure}[t]
    \centering    \includegraphics[width=\linewidth,height= 4cm]{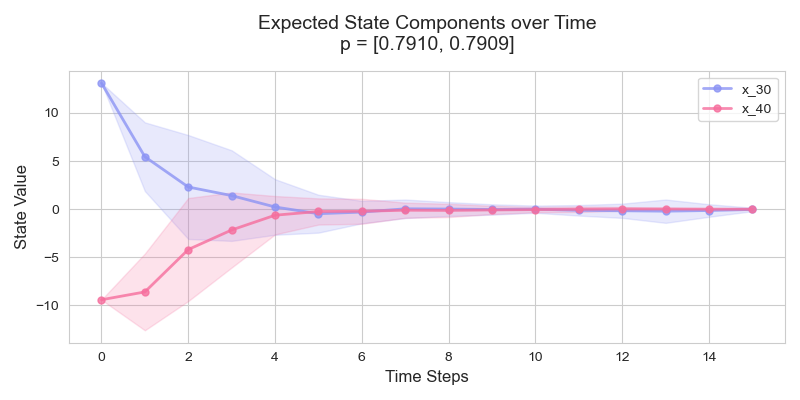}
    \caption{The stability of components $30$ and $40$ of the expected state under the adaptive sparsification strategy with parameters $0.791$ and $0.7909$ successively. Solid lines represent the mean values, and the $±$ standard deviations are shown as shaded contours.}
    \label{fig:example2_3_A_A}
\end{figure}
\begin{figure*}[t]
\centering
    \begin{subfigure}[t]{0.48\textwidth}   \includegraphics[width=\linewidth, height=4cm]{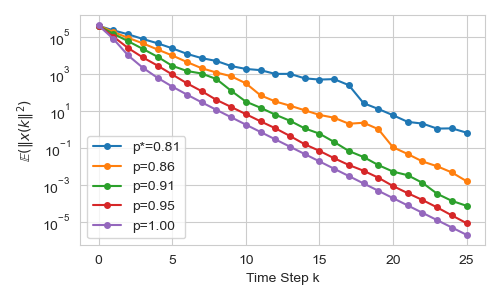}
    \caption{ The system's AMSS is satisfied when the sparsification parameter meets or exceeds  $p^\star$.}
    \label{fig:example2_4_A}
    \end{subfigure}
    \hfill
    \begin{subfigure}[t]{0.48\textwidth}
        \centering       \includegraphics[width=\linewidth,height=4cm]{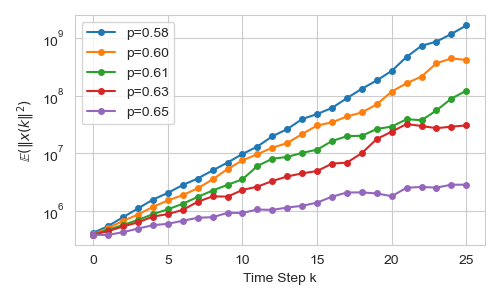}
    \caption{ The system's AMSS is not fulfilled for values of the sparsification parameter strictly lower than $p^\star$.}
    \label{fig:example2_4_div_log}
    \end{subfigure}
    \caption{The variation of $\E_{x(0),\C}\left(\norm{x(k)}^2\right)$, of the example of interconnected systems.}
    \label{fig:interconnected_systems_A}
\end{figure*}
\section{Appendix B}
In this section, we first present simulation results for a new case study involving a spatially distributed dynamical system. We then provide additional experiments for the second numerical example in the paper (large-scale power system).
\subsection{Example of interconnected systems}
We consider a discrete-time control system consisting of \( N = 20 \) spatially interconnected local subsystems, following the framework presented in \cite{haber2012identification}. The global state and input vectors are \( \mathbf{x}(k) \in \mathbb{R}^{2N} \) and \( \mathbf{u}(k) \in \mathbb{R}^{N} \), respectively. The global system dynamics are governed by:
\[
\mathbf{x}(k+1) = A \mathbf{x}(k) + B \mathbf{u}(k),
\]
where the global system matrices \( A\in \R^{2N\times 2N} \) and \(B\in \R^{2N\times N} \) are block-structured to capture subsystem interconnections:
\[
A = \begin{bmatrix}
\underline{A} & E & & \\
E & \underline{A} & E & \\
& \ddots & \ddots & \ddots \\
& & E & \underline{A} & E \\
& & & E & \underline{A}
\end{bmatrix}, \quad B = \begin{bmatrix}
\underline{B} & & \\
& \ddots & \\
& & \underline{B}
\end{bmatrix}.
\]
The local subsystem matrices $\underline{A}$, $E$ and $\underline{B}$ are given by
\[
\underline{A} = \begin{bmatrix} 1 & 0.890 \\ 0.890 & 1 \end{bmatrix}, \quad E = \begin{bmatrix} 0.0890 & 0 \\ 0 & 0.0890 \end{bmatrix}\]
\[\underline{B} = \begin{bmatrix} 3.5600 \\ 1.7800 \end{bmatrix}.
\]
\begin{figure}[t]
    \centering
\includegraphics[width=\linewidth,height=4cm]{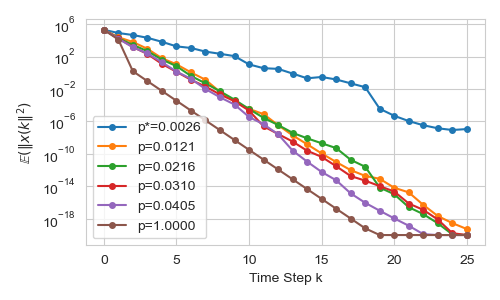}
    \caption{The variation of $\E_{x(0),\C}\left(\norm{x(k)}^2\right)$ for example of a large-scale power system with the number of nodes $n=1000$. We vary the sparsification parameter $p$ from  $ p^\star$ to 1. In this scenario, the initial condition was sampled from a Gaussian distribution with a mean of 0 and a variance of $\sigma = 100$, and the mean is estimated using Monte Carlo simulations.}
  \label{fig:example3_4_A}
\end{figure}
Let us mention that the model presented above of spatially interconnected subsystems is a powerful modeling framework for large-scale dynamical networks where individual subsystems are coupled across spatial domains. This modeling paradigm arises naturally in numerous real-world applications including automated highway systems, airplane formation flight and satellite constellations \cite{d2003distributed} and industrial settings such as cross-directional control in paper manufacturing \cite{liu2016distributed}.
\begin{figure}[t]
    \centering
\includegraphics[width=\linewidth,height = 4cm]{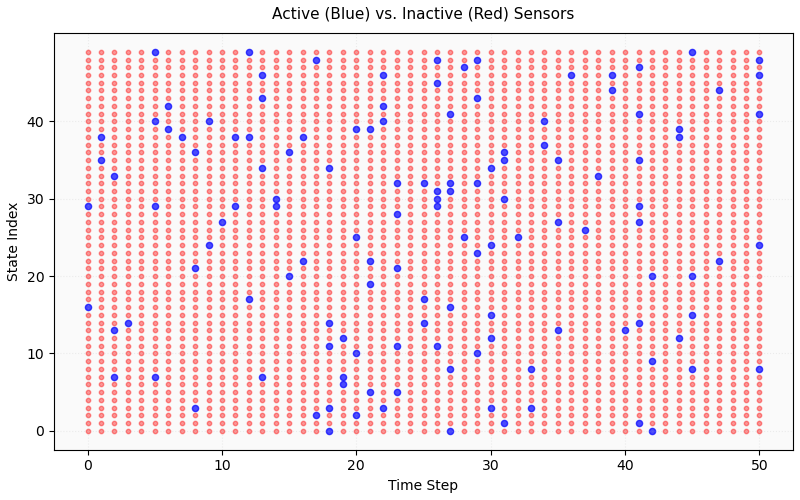}
    \caption{Active sensors versus inactive sensors at each time step for a large-scale power system where $n=50$. We use the sparsification parameter $p^\star = 0.0475$ (see table 1).}
  \label{fig:sensor_activity}
\end{figure}
\\
By applying Algorithm \ref{algo1}, we compute the Bernoulli parameter \( p^\star = 0.81 \) for the sparsification strategy, along with the corresponding control gain matrix $ K $. Figure \ref{fig:example2_1_1} illustrates the evolution of the first 20 state components, which converge to 0 over time, in agreement with the theoretical results established in Lemma \ref{expected_stability}. Furthermore, Figures \ref{fig:example2_3_A} and \ref{fig:example2_3_A_A} depict the temporal evolution of the expected states for components $1,10$ and $30,40$, respectively. Solid lines represent the mean values, while shaded contours indicate the $±$ standard deviations. In both figures, we use the sparsification parameter the sparsification vector $p^\star = [p_1^\star,\ldots,p_n^\star]$ resulting from the adaptive sparsification strategy proposed in Algorithm \ref{algo2}. Our results indicate that over time, the mean values of both components stabilize to zero, with the contours narrowing, indicating reduced variance. This convergence aligns with theoretical results in Lemma \ref{expected_stability} and Theorem \ref{mainadaptive}, suggesting that state components such as $ \{1, 10, 30, 40\} $ approach zero as time progresses.
Finally, we compare the expected square norm of the state vector for different values of parameter $p$ from $p^\star$ to $p=1$ (classical stabilization based on full state measurement without sparsification). The results shown in Fig.\ref{fig:example2_4_A} align with the theoretical result presented in Theorem  \ref{sufficient_condition}. In contrast, Fig.\ref{fig:example2_4_div_log} depicts the case where AMSS of the system is not fulfilled for some value of $p$ strictly lower than $p^\star$.
\subsection{Large-scale power system}
In this section, we provide additional illustrative figures to demonstrate the utility and effectiveness of our randomized sparsification approach, using the large power system as an example.
\\
Figure \ref{fig:example3_4_A} shows the expected square norm of the state vector for different values of parameter $p$ from $p^\star=0.0025$ to $p=1$ (classical stabilization based on full state measurement without sparsification). In this scenario, the initial condition was sampled from a Gaussian distribution with a mean of 0 and a variance of $\sigma = 100$, and the mean is estimated using Monte Carlo simulations. The results indicate that AMSS can be attained using the randomized sparsification method outlined in the theoretical findings of Theorem \ref{sufficient_condition}. Additionally, the impact of sparsification on the convergence rate is depicted: increased sparsification leads to slower convergence, consistent with the analytical insights in Remark \ref{rk}.
\\
\begin{table}[h]
\centering
\begin{tabular}{|c|c|c|c|}
\hline
\textbf{Number n} & \textbf{$p^\star$} & \textbf{$ a_n$} & \textbf{$\rho(A+BK)$} \\
\hline
50 & 0.0475 & 0.711 & 0.269\\
\hline
100 & 0.0436  & 0.677 & 0.269 \\
\hline
200 & 0.0126  & 0.719 & 0.253 \\
\hline
300 & 0.008  & 0.706 & 0.250 \\
\hline
400 & 0.006  & 0.709 & 0.250 \\
\hline
500 & 0.005 & 0.717 & 0.249 \\
\hline
600 & 0.0042&    0.717 & 0.249 \\
\hline
700 & 0.0036  & 0.72 & 0.249 \\
\hline
800 & 0.0031  & 0.71 & 0.248 \\
\hline 
900 & 0.0031  & 0.712 & 0.248 \\
\hline 
1000 & 0.0026 & 0.719 & 0.248 \\
\hline 
\end{tabular}
\caption{Parameters for different system sizes $n$. We use Algorithm 1 to find $p^\star$ and the corresponding $K$. We verify that $K\in \mathcal{K}(A,B)$.}
\label{tab:parameter}
\end{table}  
Table 1 provides a comprehensive overview of key parameters for systems of varying sizes, denoted by $n$, ranging from 50 to 1000. These parameters include the sparsification parameter $p^\star$, the largest singular value of $(I_n-B(B^\top B)^{-1}B^\top)A$ denoted by $a_n$, and the spectral radius $\rho(A + BK)$. Interestingly, the sparsification parameter decreases as the system size gets bigger, declining from $0.0475$ at $n = 50$ to $0.0026$ at $n = 1000$. Moreover, for these settings, the number of active sensors required to achieve the AMSS of the system is roughly the same, independently of the size of the system. This result confirms the utility of our randomized sparsification approach.
\\
Figure \ref{fig:sensor_activity} visualizes the number of active sensors at each time step. We consider the system with $n=50$, where the sparsification parameter $p\star = 0.0475$.  Our result demonstrates that the system's asymptotic stability can be achieved using measurements from a randomly selected subset (depicted in blue) of the state vector at each time step. Furthermore, the number of active sensors closely aligns with the expected sparsity, approximated by $ p^\star\times n.$
\bibliography{aaai2026}

\end{document}